\documentclass{ifacconf}
\usepackage{amsmath,amssymb,mathtools}
\usepackage{algorithm,algpseudocode}
\usepackage{booktabs}
\usepackage{graphicx}
\usepackage{natbib}
\usepackage[absolute]{textpos}

\begin{document}
\begin{frontmatter}

\title{A moving horizon state and parameter estimation scheme with guaranteed robust convergence\thanksref{footnoteinfo}}

\thanks[footnoteinfo]{This work was supported by the German Research Foundation (DFG) under
	the research grant MU-3929-2/1.}

\author{Julian D. Schiller},
\author{Matthias A. Müller} 
\address{Leibniz University Hannover, Institute of Automatic Control (e-mail: \{schiller,mueller\}@irt.uni-hannover.de)}

\begin{abstract}
	We propose a moving horizon estimation scheme for joint state and parameter estimation for nonlinear uncertain discrete-time systems.	
	We establish robust exponential convergence of the combined estimation error subject to process disturbances and measurement noise.
	We employ a joint incremental input/output-to-state stability ($\delta$-IOSS) Lyapunov function to characterize nonlinear detectability for the states and (constant) parameters of the system.
	Sufficient conditions for the construction of a joint $\delta$-IOSS Lyapunov function are provided for a special class of nonlinear systems using a persistence of excitation condition.
	The theoretical results are illustrated by a numerical example.
\end{abstract}


\end{frontmatter}

\begin{textblock*}{\textwidth}(1.5cm,29.3cm)
	\small{
		© 2023 the authors. This work has been accepted to IFAC for publication under a Creative Commons Licence CC-BY-NC-ND.\\
		10.1016/j.ifacol.2023.10.382
	}
\end{textblock*}

\section{Introduction}\label{sec:intro}

Robust state estimation for nonlinear systems subject to noise is a problem of high practical relevance.
However, if the underlying model is also uncertain and cannot accurately capture the real system behavior, the estimation error may even become unstable if this is not taken into account in the observer design. We address this issue by developing a robust moving horizon estimation (MHE) scheme for simultaneously estimating the states and (constant) parameters of a nonlinear system.

Recently, there have been significant developments in MHE theory, and in particular, strong robust stability properties have been established under practical conditions, see, e.g., \citep{Allan2021a,Knuefer2023,Schiller2022}.
Central to these results was the underlying detectability condition, namely $\delta$-IOSS, which is nessary and sufficient for the existence of robustly stable state estimators, cf. \citep{Allan2021,Knuefer2023}.
Procedures to verify this crucial property in practice have recently been developed in \citep{Schiller2022}.
However, the derived guarantees for MHE rely on an exact model of the real system and are therefore not valid in the case of (parametric) model uncertainties.
To address this problem, a $\min$-$\max$ MHE scheme was proposed earlier in \citep{Alessandri2012}, where at each time step a least-squares cost function is minimized for the worst case of the model uncertainties.
However, such a $\min$-$\max$ approach becomes computationally intensive for general nonlinear systems, and the worst-case consideration may be too conservative and affect estimation performance.
On the other hand, it is often beneficial not only to ensure robustness against model errors, but also to obtain an estimate of the uncertain parameters, since a good model is crucially required for, e.g., (high-performance) control, system monitoring, or fault detection.
In this context, an MHE scheme was proposed in \citep{Sui2011}, treating the parameters as additional states (with constant dynamics).
The corresponding stability analysis is based on the transformation of the extended system into an observable and an unobservable but exponentially stable subsystem, where the temporary loss of observability (due to lack of excitation) is handled by suitable regularization and adaptive weights.
However, the robustness properties have not been analyzed, and the imposed conditions (in particular, the existence of a suitable transformation) for guaranteed state and parameter convergence are not trivial to verify in practice.

An alternative approach to simultaneous state and parameter estimation is provided by \textit{adaptive observers}, the concepts of which have been extensively studied in the literature, see, e.g., the book by \cite{Ioannou2012}.
Typically, this involves a detectability/observability condition on the system states and a persistence of excitation (PE) condition to establish parameter convergence.
Different system classes (usually neglecting disturbances) have been considered, e.g., linear time-varying systems \citep{Guyader2003}, Lipschitz nonlinear systems under a linear parameterization \citep{Cho1997}, and nonlinearly parameterized systems \citep{Farza2009}.
A more general class of nonlinear systems was considered in \citep{Ibrir2018}, however, requiring the verification of a large-dimensional linear matrix inequality (LMI), the feasibility of which is generally difficult to guarantee.
Many results also consider a particular nonlinear adaptive observer canonical form (with dynamics affine in the unknown parameter and the nonlinearities depending on the measured output), cf., e.g., \citep{Bastin1988,Marino2001}, where the latter also includes a robustness analysis, and \citep{Tomei2022}, where lack of PE is countered.
An adaptive sliding mode observer was proposed in~\citep{Efimov2016}, which was transferred to a more general class of systems in \citep{Franco2020}, albeit under conditions that imply certain structural restrictions.
An alternative approach applicable to general nonlinear systems is provided by the supervisory framework, cf. e.g. \cite{Meijer2021}, where, however, the desired degree of accuracy is directly related to the number of observers that have to be simulated and evaluated in parallel.

\textit{Contribution} \ \
We propose an MHE scheme for joint state and parameter estimation of general nonlinear discrete-time systems subject to process disturbances and measurement noise.
Our arguments are based on recent MHE results \citep{Schiller2022,Allan2021a}, where only state (but no parameter) estimation is considered.
In particular, we extend the concept of $\delta$-IOSS by defining a joint \mbox{$\delta$-IOSS} Lyapunov function for the system states and the (constant) parameters, which we then use as $N$-step Lyapunov function for MHE to establish robust convergence of the state and parameter estimates to their true values.
As second contribution, we derive sufficient conditions for constructing such a joint $\delta$-IOSS Lyapunov function for a certain class of nonlinear systems that are affine in the unknown parameter, which involves arguments from the adaptive observer literature, in particular, a PE condition similar to those from \cite{Guyader2003, Efimov2016, Ibrir2018}.

\textit{Notation} \ \
The set of integers is denoted by $\mathbb{I}$, the set of all integers greater than or equal to $a$ for any $a \in \mathbb{I}$ by $\mathbb{I}_{\geq a}$, and the set of integers in the interval $[a,b]$ for any $a,b\in \mathbb{I}$ with $a\le b$ by $\mathbb{I}_{[a,b]}$.
The $n \times n$ identity matrix is denoted by $I_n$ and the $n \times m$ matrix containing only zeros is denoted by $0_{n\times m}$, where we omit the indices if the dimension is unambiguous from the context.
The weighted Euclidean norm of a vector $x \in \mathbb{R}^n$ with respect to a positive definite matrix $Q=Q^\top$ is defined as $\|x\|_Q=\sqrt{x^\top Q x}$ with $\|x\|=\|x\|_{I_n}$; the minimal and maximal eigenvalues of $Q$ are denoted by $\lambda_{\min}(Q)$ and $\lambda_{\max}(Q)$, respectively.
Given two matrices $A=A^\top$ and $B=B^\top$, we write $A \succeq B$ ($A\succ B$) if $A-B$ is positive semi-definite (positive definite).
For $A,B$ positive definite, the maximum generalized eigenvalue (i.e., the largest scalar $\lambda$ satisfying $\det (A-\lambda B) = 0$) is denoted by $\lambda_{\max}(A,B)$.

\section{Problem setup}

We consider the nonlinear uncertain discrete-time system
\begin{subequations}
	\label{eq:sys}
	\begin{align}
	\label{eq:sys_1}
	x_{t+1}&=f_{\mathrm{s}}(x_t,u_t,d_t,\theta),\\
	\label{eq:sys_2}
	y_t&=h_{\mathrm{s}}(x_t,u_t,d_t,\theta),
	\end{align}
\end{subequations}
with state $x_t\in\mathbb{X}\subseteq\mathbb{R}^n$, control input $u_t\in\mathbb{U}\subseteq\mathbb{R}^m$, disturbances $d_t\in\mathbb{D}\subseteq\mathbb{R}^{q}$, noisy output $y_t\in\mathbb{Y}\subseteq\mathbb{R}^p$, and time $t\in\mathbb{I}_{\geq 0}$.
The nonlinear continuous functions $f_{\mathrm{s}}$ and $h_{\mathrm{s}}$ represent the system dynamics and the output equation, respectively, both of which depend on an unknown but constant parameter $\theta\in\Theta\subseteq\mathbb{R}^o$.

The overall goal is to compute, at each time $t\in\mathbb{I}_{\geq 0}$, the estimates $\hat{x}_t$ and $\hat{\theta}_t$ of the current state $x_t$ and the real parameter $\theta$.
We provide conditions under which the combined estimation error
\begin{equation}\label{eq:e_def}
	e_t=\begin{bmatrix} e_{x,t} \\ e_{\theta,t} \end{bmatrix} \ \ \text{with} \ \
	e_{x,t} = x_t-\hat{x}_t, \ e_{\theta,t} = \theta-\hat{\theta}_t
\end{equation}
robustly exponentially converges to a neighborhood around the origin, i.e., that there exist constants $c_1,c_2>0$ and $\rho_1,\rho_2\in[0,1)$ such that
	\begin{equation}\label{eq:RGES}
	\|e_t\|\leq \max \left\{ c_1\rho_1^t\|e_0\|, \max_{j\in\mathbb{I}_{[1,t]}} c_2\rho_2^{j-1} \|d_{t-j}\| \right\}
	\end{equation}
for all $t \in\mathbb{I}_{\geq 0}$.
Note that this directly implies $e_t\rightarrow0$ if $d_t \rightarrow 0$ for $t\rightarrow\infty$.
To this end, a suitable detectability property of system~\eqref{eq:sys} is required.
As discussed in Section~\ref{sec:intro}, the use of the $\delta$-IOSS concept for this purpose has been central to recent advances in the field of nonlinear MHE for state estimation, (cf., e.g., \citep{Schiller2022,Allan2021a,Knuefer2023,Allan2021}).
Therefore, it seems natural to extend this notion by considering both the states and the parameters of system trajectories via the following Lyapunov function definition.

\begin{defn}[Joint $\delta$-IOSS Lyapunov function]\label{def:dIOSS}
	A function $U:\mathbb{I}_{\geq0} \times \mathbb{R}^n \times \mathbb{R}^n \times \mathbb{R}^o \times \mathbb{R}^o\rightarrow\mathbb{R}_{\geq0}$ is a (quadratically bounded) $\delta$-IOSS Lyapunov function for the system~\eqref{eq:sys} on some set $\mathbb{Z}_{{T}}$ if there exist a constant $T\in\mathbb{I}_{\geq 1}$, matrices $\underline{M},\overline{M}\succ0$, $Q,R\succeq0$, and a contraction rate $\lambda\in[0,1)$ such that for all trajectory pairs
	\begin{equation}\label{eq:cond_Z}
		\left((x_t,u_t,d_t,\theta)_{t=0}^{K-1},(\tilde{x}_t,u_t,\tilde{d}_t,\tilde{\theta})_{t=0}^{K-1}\right) \in\mathbb{Z}_T
	\end{equation}
	with $x_{t+1} = f_{\mathrm{s}}(x_t,u_t,d_t,\theta)$ and $\tilde{x}_{t+1} = f_{\mathrm{s}}(\tilde{x}_t,u_t,\tilde{d}_t,\tilde{\theta})$ for all $t\in\mathbb{I}_{[0,K-1]}$ and any $K\in\mathbb{I}_{\geq T}$, the following is satisfied:
	\begin{subequations}
		\label{eq:U}
		\begin{align}
			&\left\|\begin{bmatrix} x_t-\tilde{x}_t\\ \theta-\tilde{\theta} \end{bmatrix}\right\|_{\underline{M}}^2
			\leq U(t,x_t,\tilde{x}_t,\theta,\tilde{\theta}) \leq 
			\left\|\begin{bmatrix} x_t-\tilde{x}_t\\ \theta-\tilde{\theta} \end{bmatrix}\right\|_{\overline{M}}^2, \nonumber\\
			&\hspace{6cm} t\in\mathbb{I}_{[0,K]}, \label{eq:U_bound}\\
			&\hspace{3.2ex}U(t+1,x_{t+1},\tilde{x}_{t+1},\theta,\tilde{\theta})\nonumber\\
			&\leq  \lambda U(t,x_t,\tilde{x}_t,\theta,\tilde{\theta}) + \|d_t-\tilde{d}_t\|_Q^2 + \|y_t-\tilde{y}_t\|_R^2, \nonumber\\
			&\hspace{6cm} t\in\mathbb{I}_{[0,K-1]},\label{eq:U_dissip}
		\end{align}
	\end{subequations}
	where $y_t = h_{\mathrm{s}}(x_t,u_t,d_t,\theta)$ and $\tilde{y}_t = h_{\mathrm{s}}(\tilde{x}_t,u_t,\tilde{d}_t,\tilde{\theta})$.
\end{defn}

It is intuitively clear that requiring the $\delta$-IOSS Lyapunov function $U$ to satisfy the conditions~\eqref{eq:U} for arbitrary system trajectories is not reasonable in practice.
Therefore, we consider only certain pairs of trajectories of at least length $T$ for which~\eqref{eq:U} is satisfied, contained in some set $\mathbb{Z}_T$; later we will define~$\mathbb{Z}_T$ by means of a certain PE condition.
Such a condition, the satisfaction of which usually requires the choice of a suitable input sequence $\{u_{t}\}_{t=0}^{K-1}$, can then be mapped into the Lyapunov function due to the fact that we impose an additional explicit dependence of $U$ on time, which is in contrast to the time-invariant definition typically used in the MHE literature (cf., e.g., \citep{Schiller2022,Allan2021a}).
In the following, we use such a joint $\delta$-IOSS Lyapunov function to design an MHE scheme for joint state and parameter estimation that guarantees robust exponential convergence of the combined estimation error~\eqref{eq:e_def} as characterized in~\eqref{eq:RGES}.

\section{Moving horizon joint state and parameter estimation}

At each time $t\in\mathbb{I}_{\geq 0}$, the proposed MHE scheme considers the available input and output sequences of the system~\eqref{eq:sys} within a moving horizon of length $N_t = \min\{t, N\}$ for some $N\in\mathbb{I}_{\geq0}$. The current state and parameter estimates are obtained by solving the following nonlinear program:
\begin{subequations}\label{eq:MHE}
	\begin{align}\label{eq:MHE_0}
	&\min_{\hat{x}_{t-N_t|t},\hat{\theta}_{|t},\hat{d}_{\cdot|t}} J_t(\hat{x}_{t-N_t|t},\hat{\theta}_{|t},\hat{d}_{\cdot|t}) \\ 
	&\hspace{0.5cm} \text{s.t. }
	\hat{x}_{j+1|t}=f_{\mathrm{s}}(\hat{x}_{j|t},u_j,\hat{d}_{j|t},\hat{\theta}_{|t}), \ j\in\mathbb{I}_{[t-N_t,t-1]}, \label{eq:MHE_1} \\	
	&\hspace{1.1cm} \hat{y}_{j|t}=h_{\mathrm{s}}(\hat{x}_{j|t},u_j,\hat{d}_{j|t},\hat{\theta}_{|t}) \, j\in\mathbb{I}_{[t-N_t,t-1]}, \label{eq:MHE_2} \\	
	&\hspace{1.1cm}\hat{x}_{j|t}\in\mathbb{X}, j\in\mathbb{I}_{[t-N_t,t]}, \ \hat{\theta}_{|t}\in{\Theta}, \label{eq:MHE_3}\\
	&\hspace{1.1cm}\hat{d}_{j|t} \in\mathbb{D}, j\in\mathbb{I}_{[t-N_t,t-1]}. \label{eq:MHE_4}
	\end{align}
\end{subequations}
The decision variables $\hat{x}_{t-N_t|t}$, $\hat{\theta}_{|t}$, and $\hat{d}_{\cdot|t} = \{\hat{d}_{j|t}\}_{j=t-N_t}^{t-1}$ denote the current estimates of the state at the beginning of the horizon, the parameter, and the disturbance sequence over the horizon, respectively, estimated~at time~$t$.
Given the past input sequence $\{u_j\}_{j=t-N_t}^{t-1}$ applied to system~\eqref{eq:sys}, these decision variables (uniquely) define a sequence of state estimates $\{\hat{x}_{j|t}\}_{j=t-N_t}^{t}$ under \eqref{eq:MHE_1}.
Assuming the existence of a joint $\delta$-IOSS Lyapunov function in the sense of Definition~\ref{def:dIOSS}, we consider the cost function
\begin{align}
	&J_t(\hat{x}_{t-N_t|t},\hat{\theta}_{|t},\hat{d}_{\cdot|t}) = 2\lambda^{N_t}\left\|\begin{bmatrix} \hat{x}_{t-N_t|t}-\hat{x}_{t-N_t}\\ \hat{\theta}_{|t}-\hat{\theta}_{t-N_t} \end{bmatrix}\right\|^2_{\Gamma} \nonumber \\
	& \quad +\sum_{j=1}^{N_t}\lambda^{j-1}\left(2\|\hat{d}_{t-j|t}\|_Q^2+\|\hat{y}_{t-j|t}-y_{t-j}\|_R^2 \right), \label{eq:MHE_objective}
\end{align}
where $\hat{x}_{t-N_t}$ and $\hat{\theta}_{t-N_t}$ are the estimates obtained $N_t$ steps in the past, $\{y_j\}_{j=t-N_t}^{t-1}$ is the (measured) output sequence of system~\eqref{eq:sys}, the parameters $\lambda,Q,R$ are from the $\delta$-IOSS Lyapunov function (cf.~Definition~\ref{def:dIOSS}), and $\Gamma$ is a certain weighting matrix specified below.
We denote a minimizer to~\eqref{eq:MHE} and \eqref{eq:MHE_objective} by $(\hat{x}^*_{t-N_t|t},\hat{\theta}^*_{|t},\hat{d}_{\cdot|t}^*)$, and the corresponding optimal state sequence by $\{\hat{x}^*_{j|t}\}_{j=t-N_t}^t$. The resulting state and parameter estimates at time $t\in\mathbb{I}_{\geq0}$ are then given by $\hat{x}_t{\;=\;}\hat{x}^*_{t|t}$ and $\hat{\theta}_t{\;=\;}\hat{\theta}^*_{|t}$, respectively.
The following theorem provides an $N$-step Lyapunov function for the combined estimation error~\eqref{eq:e_def}, from which robust exponential convergence can be directly deduced.

\begin{thm}\label{thm:MHE}
	Suppose that the system~\eqref{eq:sys} admits a joint \mbox{$\delta$-IOSS} Lyapunov function $U$ according to Definition~\ref{def:dIOSS} on some $\mathbb{Z}_T$.
	Let $\Gamma=\overline{M}$ in~\eqref{eq:MHE_objective} and choose $N\in\mathbb{I}_{\geq0}$ large enough such that $N\geq T$ and \mbox{$\rho:=4\lambda_{\max}(\overline{M},\underline{M})\lambda^N<1$}.
	Suppose that for each $t\in\mathbb{I}_{\geq N}$, the real and the currently estimated trajectory form a pair satisfying
	\begin{equation}
		\left((x_j,u_j,d_j,\theta)_{j=t-N}^{t-1}, (\hat{x}^*_{j|t},u_j,\hat{d}^*_{j|t},\hat{\theta}^*_{|t})_{j=t-N}^{t-1}\right) \in\mathbb{Z}_{T}. \label{eq:cond_Z_MHE}
	\end{equation}
	Then, $W_N(x,\hat{x},\theta,\hat{\theta}) := U(N,x,\hat{x},\theta,\hat{\theta})$ is a joint $N$-step Lyapunov function on $\mathbb{Z}_T$ for the state and parameter estimation error $e_t$~\eqref{eq:e_def} satisfying
	\begin{align}
		W_N(x_t,\hat{x}_t,\theta,\hat{\theta}_t)\leq&\ \rho^N W_N(x_{t-N},\hat{x}_{t-N},\theta,\hat{\theta}_{t-N})\nonumber\\ &+ 4\sum_{j=1}^{N}\lambda^{j-1}\|d_{t-j}\|^2_Q \label{eq:Nstep_Lyap}
	\end{align}
	for all $t\in\mathbb{I}_{\geq 2N}$.
\end{thm}

\begin{pf}
	Since~\eqref{eq:cond_Z_MHE} applies, we can evaluate the $\delta$-IOSS Lyapunov function $U$ along the pair formed by the real and the estimated trajectory on the estimation horizon.
	The rest of the proof follows by suitably adapting the arguments from the proof of~\cite[Th.~1]{Schiller2022}, where we exploit the relation between the specific structure (and parameterization) of the cost function~\eqref{eq:MHE_objective} and the $\delta$-IOSS Lyapunov function $U$ satisfying~\eqref{eq:U}.
\end{pf}

\begin{rem}[Convergence]\label{rem:RGES}
	The joint $N$-step Lyapunov function~\eqref{eq:Nstep_Lyap} is only valid for all $t\in\mathbb{I}_{\geq 2N}$ due to the definition of $W_N$.
	However, if for $t\in\mathbb{I}_{[0,N-1]}$ the bounds~\eqref{eq:U} also hold for the pairs of trajectories of length $t<N$ (which generally is the case for causal---i.e., non-anticipating---$U$ via the initialization at time zero, compare the conditions of Theorem~\ref{thm:U}~below), then one can straightforwardly deduce that $e_t$~\eqref{eq:e_def} satisfies~\eqref{eq:RGES} for all $t\in\mathbb{I}_{\geq0}$ with $c_1 = 4\sqrt{\frac{\lambda_{\max}(\overline{M})}{\lambda_{\min}(\underline{M})}}$,
	$c_2 = \frac{4}{1-\sqrt[4]{\rho}}\sqrt{\frac{\lambda_{\max}(Q)}{\lambda_{\min}(\underline{M})}}$,
	$\rho_1 = \sqrt{\rho}$, and $\rho_2 = \sqrt[4]{\rho}$ by using similar arguments that were applied in the proofs of Theorem~\ref{thm:MHE} and \cite[Cor.~1]{Schiller2022}.
\end{rem}

From a theoretical point of view, Theorem~\ref{thm:MHE} is not surprising, as it is a fairly straightforward extension of recent results from the MHE context (in particular \citep{Schiller2022}, where only state (but no parameter) estimation was considered).
The underlying problem is rather the construction of \mbox{$\delta$-IOSS} Lyapunov functions satisfying Definition~\ref{def:dIOSS}; choosing $U$ quadratic with respect to a constant positive definite weighting matrix, for example, which would be a first naive attempt, is in general not feasible in~\eqref{eq:U_dissip}.
In the following section, we provide an alternative approach to the construction of a time-varying quadratic $\delta$-IOSS Lyapunov function for a particular class of nonlinear systems using arguments from the adaptive observer literature.

\section{Sufficient conditions for the construction of a joint $\delta$-IOSS Lyapunov function}
\label{sec:dIOSS}
Throughout the following, we restrict ourselves to the special case where $f$ is affine in $\theta$ and subject to additive disturbances $d$ and with $h$ linear, that is, we consider
\begin{subequations}
	\begin{align}
	f_{\mathrm{s}}(x_t,u_t,d_t,\theta) &= f(x_t,u_t) + G(x_t,u_t)\theta + Ed_t,\label{eq:sys_special_1}\\
	h_{\mathrm{s}}(x_t,u_t,d_t,\theta) &= Cx_t + Fd_t\label{eq:sys_special_2}
	\end{align}
\end{subequations}
with constant matrices $E,C,F$ of appropriate dimensions and with $G$ having the following properties.

\begin{assum}[Boundedness]\label{ass:G}
	The function $G$ is continuously differentiable and the sets $\mathbb{X}$, $\mathbb{U}$, and $\Theta$ are compact.
\end{assum}

Let us consider two trajectories of system~\eqref{eq:sys} given by $(x_{t},u_t,d_t,\theta)_{t=0}^{K-1}$ and $(\tilde{x}_t,u_t,\tilde{d}_t,\tilde{\theta})_{t=0}^{K-1}$ for any $K\in\mathbb{I}_{\geq 1}$.
Using the mean value theorem, we first note that
\begin{equation} \label{eq:MVT}
	f(x,u)-f(\tilde{x},u) = A(x,\tilde{x},u)(x-\tilde{x})
\end{equation}
with
\begin{equation}
	A(x,\tilde{x},u) := \int_{0}^{1}\frac{\partial f}{\partial x}(\tilde{x}+s({x}-\tilde{x}),u)ds \label{eq:A_def}
\end{equation}
for all $x,\tilde{x}\in\mathbb{X}$ and $u\in\mathbb{U}$.

\begin{assum}[Detectability]\label{ass:Lyap}
	There exists a continuous mapping $L:\mathbb{R}^n \times \mathbb{R}^n \times \mathbb{R}^m \rightarrow \mathbb{R}^p$ such that the matrix
	\begin{equation}\label{eq:phi}
		\Phi(x,\tilde{x},u) = A(x,\tilde{x},u)+L(x,\tilde{x},u)C
	\end{equation}
	satisfies
	\begin{equation}\label{eq:Lyap}
		\Phi(x,\tilde{x},u)^\top P\Phi(x,\tilde{x},u) \preceq \mu P
	\end{equation}
	for some $P\succ0$ and $\mu\in[0,1)$ uniformly for all $x,\tilde{x}\in\mathbb{X}$ and $u\in\mathbb{U}$.
\end{assum}

\begin{rem}[Computation of $P,L$]\label{rem:Lyap}
	The matrices $P$ and $L$ can be easily computed by transforming~\eqref{eq:Lyap} into an LMI condition using standard linear algebra and a suitable verification (e.g., using sum-of-squares (SOS) optimization).
	The additional degree of freedom provided by allowing $L$ to depend on both the states $x$ and $\tilde{x}$ can be used, e.g., to compensate for nonlinear terms in~\eqref{eq:A_def}, which is beneficial in several aspects, compare Remark~\ref{rem:online}.
	We point out that this is generally not possible in the context of (adaptive) observers due to the fact that $L$ is crucially required for performing the observer recursions at each time step and therefore enforced to be constant as in, e.g., \citep{Ibrir2018}.
\end{rem}

Having set the assumptions regarding the system state, we also need assumptions regarding the parameters. To this end, note that for any fixed $\theta$, the definition $G_\theta'(x,u) := G(x,u){\theta}$ represents a continuously differentiable function $G'_\theta: \mathbb{R}^n\times\mathbb{R}^m\rightarrow \mathbb{R}^n$.
Therefore, by the mean value theorem,
\begin{align}
	(G(x,u)-G(\tilde{x},u)){\theta} &= G'_\theta(x,u)-G'_\theta(\tilde{x},u)\nonumber \\
	& = \mathcal{G}_\theta(x,\tilde{x},u)(x-\tilde{x}), \label{eq:MVT_G}
\end{align}
where
\begin{equation}
	\mathcal{G}_\theta(x,\tilde{x},u) := \int_{0}^{1}\frac{\partial G'_{\theta}}{\partial x}(\tilde{x}+s(x-\tilde{x}),u)ds \label{eq:G_def}
\end{equation}
for all $x,\tilde{x}\in\mathbb{X}$ and $u\in\mathbb{U}$.

\begin{assum}\label{ass:CTHC}
	There exists a matrix $H$ such that
	\begin{equation}\label{eq:CTHC}
		{\mathcal{G}}_\theta(x,\tilde{x},u)^\top P {\mathcal{G}}_\theta(x,\tilde{x},u) \preceq C^\top H C
	\end{equation}
	uniformly for all $x,\tilde{x}\in\mathbb{X}$, $u\in\mathbb{U}$, and $\theta\in{\Theta}$ with $P$ from Assumption~\ref{ass:Lyap}.
\end{assum}

\begin{rem}[Conditions on $G$]\label{rem:conditions_G}
	Condition~\eqref{eq:CTHC} is linear in~$H$ and thus can be easily verified using standard LMI methods.
	We point out that Assumption~\ref{ass:CTHC} is related to the condition used in \citep{Cho1997}; furthermore, $H$ always exists for the special case where $G(x,u) = G(Cx,u)$, which includes the important classes of nonlinear adaptive observer canonical forms that are often considered in the adaptive observer literature (cf., e.g., \citep{Tomei2022,Marino2001}).
\end{rem}

We define the recursions
\begin{align}
	Y_{t+1} &= \Phi(x_t,\tilde{x}_t,u_t) Y_t + G(\tilde{x}_t,u_t), \  t\in\mathbb{I}_{[0,K-1]}\label{eq:Y_def}\\
	S_{t+1} &= \eta S_t + Y_t^\top C^\top C Y_t, \  t\in\mathbb{I}_{[0,K-1]}\label{eq:S_def}
\end{align}
for some fixed $Y_0\in\mathbb{R}^{n \times o}$, $S_0\in\mathbb{R}^{o\times o}$ and $\eta\in(0,1)$.
As usual in the context of system identification/parameter estimation, we need a PE condition in order to guarantee exponential convergence of the parameter estimation error, which in our case characterizes the set $\mathbb{Z}_T$ from Definition~\ref{def:dIOSS}.
More precisely, for some fixed $T\in\mathbb{I}_{\geq1},\alpha>0$, we define
\begin{align}
	\mathbb{Z}_T := \Big\{&\left((x_t,u_t,d_t,\theta)_{t=0}^{K-1},(\tilde{x}_t,u_t,\tilde{d}_t,\tilde{\theta})_{t=0}^{K-1}\right)\nonumber \\
	& \in\left(\mathbb{X}^K\times\mathbb{U}^K\times\mathbb{D}^K\times\Theta\right)^2: K\in\mathbb{I}_{\geq T}, \nonumber\\
	& x_{t+1} = f_{\mathrm{s}}(x_t,u_t,d_t,\theta), \ t \in \mathbb{I}_{[0,K-1]}, \nonumber\\
	& \tilde{x}_{t+1} = f_{\mathrm{s}}(\tilde{x}_t,u_t,\tilde{d}_t,\tilde{\theta}),\ t \in \mathbb{I}_{[0,K-1]}, \nonumber \\
	& \sum_{j=1}^{T} Y_{t-j}^\top C^\top CY_{t-j} \succeq \alpha I_o, \ t \in \mathbb{I}_{[T,K]} \Big\}. \label{eq:Z_def}
\end{align}

Our PE condition is essentially similar to those appearing in the adaptive observer literature (cf., e.g., \citep{Guyader2003, Efimov2016, Ibrir2018}), albeit we enforce it by considering only trajectory pairs belonging to $\mathbb{Z}_T$.
Note that the \textit{a priori} verification for the general case is still an open problem; instead, membership of trajectory pairs to the set $\mathbb{Z}_T$ can be verified by simulations, or---in special cases---online, compare Remark~\ref{rem:online}.
A trajectory pair in $\mathbb{Z}_T$ directly implies positive definiteness and uniform boundedness of $S_t$ as the following lemma shows.

\begin{lem}\label{lem:PE}
	Let some $Y_0$ and $S_0 \succ \alpha I_o$ be given.
	Then, there exist $\sigma_1,\sigma_2>0$ such that
	\begin{equation}\label{eq:lem_PE}
		\sigma_1I_o \preceq S_t \preceq \sigma_2I_o
	\end{equation}
	 for all $t\in\mathbb{I}_{[0,K]}$ and all trajectory pairs satisfying~\eqref{eq:cond_Z} with $\mathbb{Z}_T$ from~\eqref{eq:Z_def} for any $K\in\mathbb{I}_{\geq T}$.
\end{lem}

The proof is shifted to the appendix. We are now in the position to state the following result.

\begin{thm}\label{thm:U}
	Let Assumptions~\ref{ass:G}, \ref{ass:Lyap}, and \ref{ass:CTHC} hold and let
	\begin{equation}\label{eq:thm_U}
		U(t,x_t,\tilde{x}_t,\theta,\tilde{\theta}) := 
		\left\|\begin{bmatrix}x_t-\tilde{x}_t\\\theta-\tilde{\theta}\end{bmatrix}\right\|^2_
		{M_t}, \ t\in\mathbb{I}_{[0,K]}
	\end{equation}
	with
	\begin{equation}\label{eq:M_def}
		M_t := \begin{bmatrix} P & -PY_t \\ -{Y_t^{\top}} P & \ \ Y_t^{\top} P Y_t + aS_t \end{bmatrix}, \ t\in\mathbb{I}_{[0,K]}.
	\end{equation}
	Then, there exists some $a>0$ small enough such that $U$~\eqref{eq:thm_U} is a joint $\delta$-IOSS Lyapunov function on $\mathbb{Z}_T$ according to Definition~\ref{def:dIOSS}, i.e., \eqref{eq:U} is satisfied for all trajectory pairs satisfying~\eqref{eq:cond_Z} with $\mathbb{Z}_T$ from~\eqref{eq:Z_def} for any $K\in\mathbb{I}_{\geq T}$.
\end{thm}

\begin{pf}
	Inspired by adaptive observer literature (compare, e.g., \citep{Guyader2003,Bastin1988}), we employ the (filtered) transformation
	\begin{equation}\label{eq:z_def}
		\begin{bmatrix} z_t\\ \theta \end{bmatrix} 
		= T^{(z)}_t \begin{bmatrix} x_t\\ \theta \end{bmatrix}, \ \
		\begin{bmatrix} \tilde{z}_t\\ \tilde{\theta} \end{bmatrix} 
		= T^{(z)}_t
		\begin{bmatrix} \tilde{x}_t\\ \tilde{\theta} \end{bmatrix}, \ \ T^{(z)}_t = \begin{pmatrix} I & -Y_t \\ 0 & I\end{pmatrix}.
	\end{equation}
	Note that the time-varying transformation matrix $T_t^{(z)}$ is non-singular independent of $t$ and $K$.
	Using~\eqref{eq:M_def} and \eqref{eq:z_def}, observe that $U$ in~\eqref{eq:thm_U} can now be equivalently re-written as
	\begin{equation}
		U(t,x_t,\tilde{x}_t,\theta,\tilde{\theta}) = \|z_t-\tilde{z}_t\|_P^2 + a\|\theta-\tilde{\theta}\|_{S_t}^2.\label{eq:U_def}
	\end{equation}
	Define $W(z_t,\tilde{z}_t) := \|z_t-\tilde{z}_t\|_P^2$ and $V(t,\theta,\tilde{\theta}) = \|\theta-\tilde{\theta}\|_{S_t}^2$.	
	The remaining proof is structured in three parts: we first establish Lyapunov-like properties for $W$, second for $V$, and finally for $U$.
	
	\textbf{Part I.} 
	By the dynamics~\eqref{eq:sys_1} with \eqref{eq:sys_special_1}, the difference $z_t-\tilde{z}_t$ evolves according to
	\begin{align*}
		&\ z_{t+1}-\tilde{z}_{t+1} = x_{t+1}-\tilde{x}_{t+1}-Y_{t+1}(\theta-\tilde{\theta})\\
		\stackrel{\eqref{eq:sys_special_1}, \eqref{eq:Y_def}}{=}&\ f(x_t,u_t)-f(\tilde{x}_t,u_t) -\Phi(x_t,\tilde{x}_t,u_t) Y_t(\theta-\tilde{\theta})  \\
		& \quad + \Delta_{G}(x_t,\tilde{x},u_t,\theta) +E(d_t-\tilde{d}_t),
	\end{align*}
	where $\Delta_G(x_t,\tilde{x}_t,u_t,\theta) := (G(x_t,u_t)- {G}(\tilde{x}_t,u_t))\theta$.
	Adding $0=L(x_t,\tilde{x}_t,u_t)(y_t-\tilde{y}_t-(y_t-\tilde{y}_t))$ with ${y}_t=C{x}_t+F{d}_t$ and $\tilde{y}_t=C\tilde{x}_t+F\tilde{d}_t$ by \eqref{eq:sys_2} and \eqref{eq:sys_special_2} to the right-hand side together with application of \eqref{eq:MVT} and \eqref{eq:phi} yields
	\begin{align*}%
		&\ z_{t+1}-\tilde{z}_{t+1} \\
		=&\ \Phi(x_t,\tilde{x}_t,u_t)(z_t-\tilde{z}_t) + (E+L(x_t,\tilde{x}_t,u_t)F)(d_t-\tilde{d}_t)\\
		& + \Delta_{G}(x_t,\tilde{x}_t,u_t,\theta) - L(x_t,\tilde{x}_t,u_t)(y_t-\tilde{y}_t).
	\end{align*}
	Using the Cauchy-Schwarz inequality together with fact that $(a_1+a_2)^2\leq(1+\epsilon)a_1^2 + \frac{1+\epsilon}{\epsilon}a_2^2$ for any $\epsilon>0$, $a_1,a_2\geq0$, we obtain that
	\begin{align}
		&\ W(z_{t+1},\tilde{z}_{t+1})\nonumber\\
		\leq&\ (1+\epsilon_1)\|\Phi(x_t,\tilde{x}_t,u_t)(z_t-\tilde{z}_t)\|^2_P \nonumber\\
		& + \frac{3(1+\epsilon_1)}{\epsilon_1}\big(\|(E+L(x_t,\tilde{x}_t,u_t)F)(d_t-\tilde{d}_t)\|^2_P\label{eq:proof_W}\\
		& + \|\Delta_{G}(x_t,\tilde{x}_t,u_t,\theta)\|^2_P  + \|L(x_t,\tilde{x}_t,u_t)(y_t-\tilde{y}_t)\|_P^2\big). \nonumber
	\end{align}
	By Assumption~\ref{ass:Lyap}, we note that
	\begin{equation}
		\|\Phi(x_t,\tilde{x}_t,u_t)(z_t-\tilde{z}_t)\|_P^2 \leq \mu W(z_t,\tilde{z}_t).\label{eq:proof_W1}
	\end{equation}
	From Assumption~\ref{ass:CTHC}, we can infer that
	\begin{align}
		\|\Delta_{G}(x_t{,}\tilde{x}_t{,}u_t,\theta)\|^2_P {\,\leq\,} 2\|y_t{\,-\,}\tilde{y}_t\|_H^2 {\,+\,} 2\|F(d_t{\,-\,}\tilde{d}_t)\|_H^2 \label{eq:proof_W2}
	\end{align}
	by using \eqref{eq:MVT_G}, \eqref{eq:CTHC}, and \eqref{eq:sys_special_2}.
	Application of~\eqref{eq:proof_W1} and~\eqref{eq:proof_W2} to \eqref{eq:proof_W} then yields
	\begin{align}
		&\ W(z_{t+1},\tilde{z}_{t+1})\nonumber\\
		\leq&\ (1+\epsilon_1)\mu W(z_t) + \frac{3(1+\epsilon_1)}{\epsilon_1}\Big(2\|F(d_t-\tilde{d}_t)\|_H^2 \nonumber\\
		&\ + \|(E+L(x_t,\tilde{x}_t,u_t)F)(d_t-\tilde{d}_t)\|^2_P \nonumber\\
		&\ + \|L(x_t,\tilde{x}_t,u_t)(y_t-\tilde{y}_t)\|_P^2 + 2\|y_t-\tilde{y}_t\|_H^2\Big). \label{eq:proof_part_I}
	\end{align}
	
	\textbf{Part II.} Now consider $V$.
	From the one-step recursion~\eqref{eq:S_def} and the definitions of $z_t$ and $\tilde{z}_t$ in \eqref{eq:z_def}, we obtain
	\begin{align}
		&\ V(t+1,\theta,\tilde{\theta}) \stackrel{\eqref{eq:S_def}}{=} \eta\|\theta-\tilde{\theta}\|^2_{S_t} + \|CY_t(\theta-\tilde{\theta})\|^2\nonumber\\
		\stackrel{\eqref{eq:z_def},\eqref{eq:sys_special_2}}{=}\hspace{-0.75ex}&\ \eta V(t,\theta,\tilde{\theta}) + \|y_t-\tilde{y}_t - C(z_t-\tilde{z}_t) -F(d_t-\tilde{d}_t)\|^2\nonumber\\
		\leq&\ \eta V(t,\theta,\tilde{\theta}) + \frac{1+\epsilon_2}{\lambda_{\min}(P)}\|C\|^2W(z_t,\tilde{z}_t) \nonumber\\
		&\quad + \frac{2(1+\epsilon_2)}{\epsilon_2}\big(\| y_t-\tilde{y}_t\|^2 + \|F(d_t-\tilde{d}_t)\|^2\big) \label{eq:proof_part_II}
	\end{align}
	for any $\epsilon_2>0$, where we used similar arguments that were applied to obtain~\eqref{eq:proof_W} together with the definition of $W$.
	Positive definiteness of $V$ follows from uniform boundedness of $S_t$, i.e., $\sigma_1\|\theta-\tilde{\theta}\|^2 \leq V(t,\theta,\tilde{\theta}) \leq \sigma_2\|\theta-\tilde{\theta}\|^2$ for $t \in \mathbb{I}_{[0,K]}$ with $\sigma_1,\sigma_2>0$ from Lemma~\ref{lem:PE} for all $K\in\mathbb{I}_{\geq T}$.
	
	\textbf{Part III.} Finally, we consider $U$~\eqref{eq:U_def}, i.e., the sum of $W$ and $aV$, and define
	\begin{align}
		\bar{\mu} :=  (1+\epsilon_1)\mu + \frac{a(1+\epsilon_2)}{\lambda_{\min}(P)}\|C\|^2. \label{eq:mu_bar}
	\end{align}
	By application of \eqref{eq:proof_part_I} and \eqref{eq:proof_part_II}, we thus obtain
	\begin{align}
		&\ U(t+1,x_{t+1},\tilde{x}_{t+1},\theta,\tilde{\theta}) \label{eq:proof_U} \\
		\leq&\ \bar{\mu}W(z_{t},\tilde{z}_{t}) + a\eta V_{t}(\theta,\tilde{\theta})  + \|d_t-\tilde{d}_t\|_{Q}^2 + \|y_t-\tilde{y}_t\|_{R}^2,\nonumber
	\end{align}
	where
	\begin{align*}
		Q:= \max_{x,\tilde{x}\in\mathbb{X},u\in\mathbb{U}} \bar{Q}(x,\tilde{x},u), \quad 
		R:= \max_{x,\tilde{x}\in\mathbb{X},u\in\mathbb{U}} \bar{R}(x,\tilde{x},u)
	\end{align*}
	with
	\begin{align*}
		\bar{Q}(x,\tilde{x},u) :=&\ a\frac{2(1+\epsilon_2)}{\epsilon_2}F^\top F + \frac{3(1+\epsilon_1)}{\epsilon_1}\big(2F^\top HF \\
		&\ +(E+L(x,\tilde{x},u)F)^\top P (E+L(x,\tilde{x},u)F)\big),\\ 
		\bar{R}(x,\tilde{x},u) :=&\ a\frac{2(1+\epsilon_2)}{\epsilon_2}I_p\nonumber \\
		&\ + \frac{3(1+\epsilon_1)}{\epsilon_1} \left(L(x,\tilde{x},u)^\top P L(x,\tilde{x},u) + 2H \right).
	\end{align*}
	Note that $Q$ and $R$ are well-defined due to compactness of $\mathbb{X},\mathbb{U}$ (Assumption~\ref{ass:G}) and continuity of $L$ (Assumption~\ref{ass:Lyap}).
	Now we can choose $\epsilon_1,\epsilon_2,a$ small enough such that $\bar{\mu}\in(0,1)$ in \eqref{eq:mu_bar} and define $\lambda := \max \{\bar{\mu},\eta\}\in(0,1)$.
	Hence, from~\eqref{eq:proof_U}, we can conclude that $U(t+1,x_{t+1},\tilde{x}_{t+1},\theta,\tilde{\theta}) \leq \lambda  U(t,x_t,\tilde{x}_t,\theta,\tilde{\theta}) + \|d_t-\tilde{d}_t\|_{Q}^2 + \|y_t-\tilde{y}_t\|_{R}^2$.
		
	It remains to show boundedness of $U$ as stated in~\eqref{eq:U_bound}.
	To this end, recall the transformation from~\eqref{eq:z_def} and \eqref{eq:U_def} and note that $\|P\|,\|S_t\|,\|T^{(z)}_t\|$ are uniformly upper and lower bounded (away from zero) for all times by Assumption~\ref{ass:Lyap}, Lemma~\ref{lem:PE}, and uniform boundedness of $Y_t$ (cf. the proof of Lemma~\ref{lem:PE} for further details), respectively. Therefore, we can conclude that there exist matrices $\underline{M},\overline{M}\succ0$ such that $\underline{M} \preceq M_t \preceq \overline{M}$ uniformly for all $t\in\mathbb{I}_{[0,K]}$ for all trajectory pairs satisfying~\eqref{eq:cond_Z} with $\mathbb{Z}_T$ from~\eqref{eq:Z_def} for any $K\in\mathbb{I}_{\geq T}$, which establishes~\eqref{eq:U_bound} and thus finishes this proof.
\end{pf}

\begin{rem}[Simplified design and online verification]\label{rem:online}
	If\\$\phi$~\eqref{eq:phi} can be made constant by a suitable choice of~$L$, the design effort of the proposed MHE scheme can be reduced considerably. Namely, we can replace the \textit{a priori} computation of $\alpha,T,\underline{M},\overline{M}$ by an online verification of certain conditions.	
	To this end, we choose some $Y_0,S_0$ such that $M_0\succ 0$ and set $\Gamma=M_0$ in \eqref{eq:MHE_objective}.
	Furthermore, we choose $N$ such that $4\kappa\lambda^N<1$ for some fixed $\kappa>0$, so that the cost function~\eqref{eq:MHE_objective} can be implemented given suitable values for $\lambda,Q,R$.
	Now for any $t\in \mathbb{I}_{\geq0}$, let $Y_{j|t}$, $S_{j|t}$, $M_{j|t}$ with $j \in \mathbb{I}_{[0,N_t]}$ represent $Y_j$, $S_j$, $M_j$ given the currently estimated trajectory $(\hat{x}^*_{t-N_t+j|t},u_{t-N_t+j},\hat{d}^*_{t-N_t+j|t},\hat{\theta}^*_{|t})_{j=0}^{N_t-1}$, i.e., determined by \eqref{eq:Y_def}, \eqref{eq:S_def}, and \eqref{eq:M_def}, respectively, with $\tilde{x}_j:=\hat{x}^*_{t-N_t+j|t}$.
	Note that $Y_{j|t}$ (and hence $S_{j|t}$ and $M_{j|t}$) can explicitly be computed, since the recursion rule~\eqref{eq:Y_def} only depends on known quantities in case of a constant $\Phi$ (in particular, it does not depend on the unknown true system state~$x_t$).
	Hence, for a given value of $\alpha>0$, we can verify the following two conditions online:
	\begin{align}
		&\lambda_{\max}(M_0,M_{N_t|t})\leq\kappa, \ \forall t\in\mathbb{I}_{\geq 0},
		\label{eq:cond_online_1}\\
		&\lambda_{\min}\Big(\sum_{j=0}^{N-1}Y_{j|t}^\top C^\top CY_{j|t}\Big)>\alpha I_{o},  \ \forall t\in\mathbb{I}_{\geq N}.
		\label{eq:cond_online_2}
	\end{align}
	Satisfaction of~\eqref{eq:cond_online_2} directly implies that condition~\eqref{eq:cond_Z_MHE} of Theorem~\ref{thm:MHE} holds with $T=N$.
	If also~\eqref{eq:cond_online_1} holds, we can replace $\lambda_{\max}(\overline{M},\underline{M})$ by $\kappa$ in the proof of~Theorem~\ref{thm:MHE} using the facts that $\Gamma=M_0$ and $U$~\eqref{eq:thm_U} is quadratic.
	Consequently, if the conditions~\eqref{eq:cond_online_1} and \eqref{eq:cond_online_2} are satisfied and $\lambda,Q,R$ are chosen appropriately, then the estimation error $e_t$~\eqref{eq:e_def} is guaranteed to robustly exponentially converge as characterized in~\eqref{eq:RGES}, compare~Remark~\ref{rem:RGES} and see also the simulation example in Section~\ref{sec:example}.
\end{rem}

\section{Numerical example}\label{sec:example}
\begin{figure}
	\centering
	\includegraphics{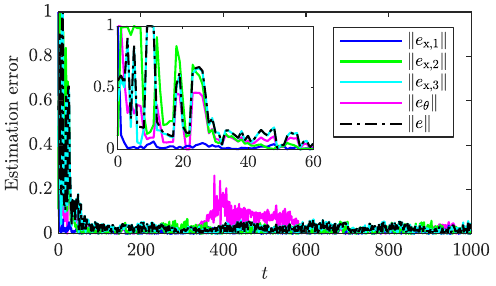}
	\caption{State and parameter estimation errors normalized to unity over time. Since conditions~\eqref{eq:cond_online_1} and \eqref{eq:cond_online_2} hold during the simulation, robust convergence of the combined estimation error $e$ (black) is guaranteed.}
	\label{fig:1}
\end{figure}

To illustrate our results, we consider the following system
\begin{align*}
	x_1^+ &= x_1 + t_{\Delta}b_1(x_2-a_1x_1-a_2x_1^2-a_3x_1^3) + d_1,\\
	x_2^+ &= x_2 + t_{\Delta}(x_1-x_2+x_3) + d_2,\\
	x_3^+ &= x_3-t_{\Delta}b_2x_2 + d_3,\\
	y &= x_1 + d_4,
\end{align*}
which corresponds to the Euler-discretized modiﬁed Chua's circuit system from~\cite{Yang2015} using the step size $t_{\Delta} = 0.01$ under additional disturbances $d\in\mathbb{R}^4$, where only $x_1$ can be measured.
The parameters are $b_1=12.8$, $b_2 = 19.1$, $a_1=0.6$, $a_2 = -1.1$, $a_3 = 0.45$, which leads to a chaotic behavior of the system.
We consider the initial condition $x_0=[2,0.1,-2]^\top$, assume that $x$ evolves in the (known) set $\mathbb{X} = [-5,5]\times[-1,1]\times[-3,3]$, and choose the naive initial estimate $\hat{x}_0=[0,0,0]^\top$.
Furthermore, we assume that the exact parameter $a_3=\theta$ is unknown but contained in the set $\Theta=[0.2,0.8]$.
In the following, we treat $d$ as a uniformly distributed random variable with $|d_i|\leq 10^{-3}, i=1,2,3$ for the process disturbance and $|d_4|\leq 0.1$ for the measurement noise.
Assumptions~\ref{ass:Lyap} and \ref{ass:CTHC} are verified in Matlab using the SOS toolbox Yalmip \citep{Loefberg2009} in combination with the semidefinite programming solver Mosek \citep{MOSEKApS2019}.
Here, we choose $L$ linear in $x$ and $\tilde{x}$ such that $\phi$ in~\eqref{eq:phi} becomes constant, leading to the contraction rate $\mu=0.9$, and we choose the parameters $a,\epsilon_1,\epsilon_2$ such that $\bar{\mu}=\eta=\lambda=0.911$. Then, we verify Assumption~\ref{ass:CTHC} on $\mathbb{X},{\Theta}$ and compute the matrices $Q,R$.
Since $\phi$ is constant, we can apply Remark~\ref{rem:online}, where we choose $N=200$ such that $4\kappa\lambda^N<1$ with $\kappa=10^7$.
Figure~\ref{fig:1} shows the estimation error of each state and the parameter as well as the combined estimation error~\eqref{eq:e_def} normalized to unity over time, revealing exponential convergence to a neighborhood around the origin as guaranteed by Theorem~\ref{thm:MHE} (using the modifications from Remark~\ref{rem:online} and the fact that conditions~\eqref{eq:cond_online_1} and \eqref{eq:cond_online_2} are satisfied during the simulation).
The relatively large parameter estimation error in the time interval $[400,600]$ is due to temporary weak excitation, which comes into play especially in combination with the measurement noise present here.
The estimation results could be easily improved by choosing $N$ larger or by temporarily stopping parameter estimation when weak excitation is detected.

\section{Conclusion}
We proposed a moving horizon estimation scheme for joint state and parameter estimation and provided sufficient conditions for guaranteed robust exponential convergence.
In particular, we used a joint $\delta$-IOSS Lyapunov function as a detectability condition for the system states and parameters to construct an $N$-step Lyapunov function for the combined state and parameter estimation error.
Moreover, we provided a systematic approach for the construction of {$\delta$-IOSS} Lyapunov functions for the special class of nonlinear systems that are affine in the parameters using a persistence of excitation condition.
Extensions to treat more general classes of nonlinear systems and to relax some of the conditions are the subject of future work.

\appendix

\section{Proof of Lemma~\ref{lem:PE}}

\begin{pf}
	We start by noting that the recursion~\eqref{eq:S_def} implies $S_{t} = \eta^{t}S_{0} + \sum_{j={1}}^{t} \eta^{j-1} Y_{t-j}^\top C^\top CY_{t-j}$, $t\in\mathbb{I}_{[0,K]}$, which directly yields the lower bound $S_t\succeq \eta^{T-1}\alpha I_o$ for $t\in\mathbb{I}_{[0,K]}$.
	For the upper bound, we obtain that $\|S_t\| \leq \eta^{t}\|S_0\| + \sum_{j=1}^{t} \eta^{j-1} \|CY_{t-j}\|^2$	for $t\in\mathbb{I}_{[0,K]}$.
	Since $Y_{t}$ satisfies \eqref{eq:Y_def}, we note that $Y_t$ is (uniformly) upper bounded for all $t\in\mathbb{I}_{\geq 0}$ due to the fact that $\Phi$ satisfies~\eqref{eq:Lyap} and $G$ is uniformly bounded using Assumption~\ref{ass:G}.
	Hence, there exists some $\beta>0$ such that $\|CY_t\|\leq \beta$ uniformly for all $t\in\mathbb{I}_{\geq0}$.
	Finally, by applying the geometric series and defining $\sigma_1 := \eta^{T-1}\alpha$ and $\sigma_2 := \big(\|S_0\| + \frac{\eta}{1-\eta} \beta^2\big)$, we derive~\eqref{eq:lem_PE}, which concludes this proof.
\end{pf}

\end{document}